\documentclass{webofc}
\usepackage[varg]{txfonts}   
\usepackage{caption}
\begin{document}
\title{Overdensity of SubMillimiter Galaxies in the GJ526 Field mapped with the NIKA2 Camera}
%
%

\author{\firstname{J.-F.}~\lastname{Lestrade}\inst{\ref{label_of_speakers_institute}}\fnsep\thanks{\email{jean-francois.lestrade@obspm.fr}}
  \and \firstname{R.}~\lastname{Adam} \inst{\ref{LLR}}
  \and  \firstname{P.}~\lastname{Ade} \inst{\ref{Cardiff}}
  \and  \firstname{H.}~\lastname{Ajeddig} \inst{\ref{CEA}}
  \and  \firstname{P.}~\lastname{Andr\'e} \inst{\ref{CEA}}
  \and \firstname{E.}~\lastname{Artis} \inst{\ref{LPSC}}
  \and  \firstname{H.}~\lastname{Aussel} \inst{\ref{CEA}}
  \and  \firstname{A.}~\lastname{Beelen} \inst{\ref{IAS}}
  \and  \firstname{A.}~\lastname{Beno\^it} \inst{\ref{Neel}}
  \and  \firstname{S.}~\lastname{Berta} \inst{\ref{IRAMF}}
  \and  \firstname{L.}~\lastname{Bing} \inst{\ref{LAM}}
  \and  \firstname{O.}~\lastname{Bourrion} \inst{\ref{LPSC}}
  \and  \firstname{M.}~\lastname{Calvo} \inst{\ref{Neel}}
  \and  \firstname{A.}~\lastname{Catalano} \inst{\ref{LPSC}}
  \and  \firstname{A.}~\lastname{Coulais} \inst{\ref{label_of_speakers_institute}}
  \and  \firstname{M.}~\lastname{De~Petris} \inst{\ref{Roma}}
  \and  \firstname{F.-X.}~\lastname{D\'esert} \inst{\ref{IPAG}}
  \and  \firstname{S.}~\lastname{Doyle} \inst{\ref{Cardiff}}
  \and  \firstname{E.~F.~C.}~\lastname{Driessen} \inst{\ref{IRAMF}}
  \and  \firstname{A.}~\lastname{Gomez} \inst{\ref{CAB}}
  \and  \firstname{J.}~\lastname{Goupy} \inst{\ref{Neel}}
  \and  \firstname{F.}~\lastname{K\'eruzor\'e} \inst{\ref{LPSC}}
  \and  \firstname{C.}~\lastname{Kramer} \inst{\ref{IRAME}}
  \and  \firstname{B.}~\lastname{Ladjelate} \inst{\ref{IRAME}}
  \and  \firstname{G.}~\lastname{Lagache} \inst{\ref{LAM}}
  \and  \firstname{S.}~\lastname{Leclercq} \inst{\ref{IRAMF}}
  \and  \firstname{J.-F.}~\lastname{Mac\'ias-P\'erez} \inst{\ref{LPSC}}
  \and  \firstname{A.}~\lastname{Maury} \inst{\ref{CEA}}
  \and  \firstname{P.}~\lastname{Mauskopf} \inst{\ref{Cardiff},\ref{Arizona}}
  \and \firstname{F.}~\lastname{Mayet} \inst{\ref{LPSC}}
  \and  \firstname{A.}~\lastname{Monfardini} \inst{\ref{Neel}}
  \and  \firstname{M.}~\lastname{Mu\~noz-Echeverr\'ia} \inst{\ref{LPSC}}
  \and  \firstname{L.}~\lastname{Perotto} \inst{\ref{LPSC}}
  \and  \firstname{G.}~\lastname{Pisano} \inst{\ref{Cardiff}}
  \and  \firstname{N.}~\lastname{Ponthieu} \inst{\ref{IPAG}}
  \and  \firstname{V.}~\lastname{Rev\'eret} \inst{\ref{CEA}}
  \and  \firstname{A.~J.}~\lastname{Rigby} \inst{\ref{Cardiff}}
  \and  \firstname{A.}~\lastname{Ritacco} \inst{\ref{IAS}, \ref{ENS}}
  \and  \firstname{C.}~\lastname{Romero} \inst{\ref{Pennsylvanie}}
  \and  \firstname{H.}~\lastname{Roussel} \inst{\ref{IAP}}
  \and  \firstname{F.}~\lastname{Ruppin} \inst{\ref{MIT}}
  \and  \firstname{K.}~\lastname{Schuster} \inst{\ref{IRAMF}}
  \and  \firstname{S.}~\lastname{Shu} \inst{\ref{Caltech}}
  \and  \firstname{A.}~\lastname{Sievers} \inst{\ref{IRAME}}
  \and  \firstname{C.}~\lastname{Tucker} \inst{\ref{Cardiff}}
  \and  \firstname{R.}~\lastname{Zylka} \inst{\ref{IRAMF}}
}
\institute{
 LERMA, Observatoire de Paris, PSL Research University, CNRS, Sorbonne Universit\'e, UPMC, 75014 Paris, France
  \label{label_of_speakers_institute}
  \and
  LLR (Laboratoire Leprince-Ringuet), CNRS, École Polytechnique, Institut Polytechnique de Paris, Palaiseau, France
  \label{LLR}
  \and
  School of Physics and Astronomy, Cardiff University, Queen’s Buildings, The Parade, Cardiff, CF24 3AA, UK 
  \label{Cardiff}
  \and
  AIM, CEA, CNRS, Universit\'e Paris-Saclay, Universit\'e Paris Diderot, Sorbonne Paris Cit\'e, 91191 Gif-sur-Yvette, France
  \label{CEA}
  \and
  Univ. Grenoble Alpes, CNRS, Grenoble INP, LPSC-IN2P3, 53, avenue des Martyrs, 38000 Grenoble, France
  \label{LPSC}
  \and
  Institut d'Astrophysique Spatiale (IAS), CNRS, Universit\'e Paris Sud, Orsay, France
  \label{IAS}
  \and
  Institut N\'eel, CNRS, Universit\'e Grenoble Alpes, France
  \label{Neel}
  \and
  Institut de RadioAstronomie Millim\'etrique (IRAM), Grenoble, France
  \label{IRAMF}
  \and
  Aix Marseille Univ, CNRS, CNES, LAM (Laboratoire d'Astrophysique de Marseille), Marseille, France
  \label{LAM}
  \and 
  Dipartimento di Fisica, Sapienza Universit\`a di Roma, Piazzale Aldo Moro 5, I-00185 Roma, Italy
  \label{Roma}
  \and
  Univ. Grenoble Alpes, CNRS, IPAG, 38000 Grenoble, France 
  \label{IPAG}
  \and
  Centro de Astrobiolog\'ia (CSIC-INTA), Torrej\'on de Ardoz, 28850 Madrid, Spain
  \label{CAB}
  \and  
  Instituto de Radioastronom\'ia Milim\'etrica (IRAM), Granada, Spain
  \label{IRAME}
  \and
  School of Earth and Space Exploration and Department of Physics, Arizona State University, Tempe, AZ 85287, USA
  \label{Arizona}
  \and 
  Laboratoire de Physique de l’\'Ecole Normale Sup\'erieure, ENS, PSL Research University, CNRS, Sorbonne Universit\'e, Universit\'e de Paris, 75005 Paris, France 
  \label{ENS}
  \and
  Department of Physics and Astronomy, University of Pennsylvania, 209 South 33rd Street, Philadelphia, PA, 19104, USA
  \label{Pennsylvanie}
  \and 
  Institut d'Astrophysique de Paris, CNRS (UMR7095), 98 bis boulevard Arago, 75014 Paris, France
  \label{IAP}
  \and 
  Kavli Institute for Astrophysics and Space Research, Massachusetts Institute of Technology, Cambridge, MA 02139, USA
  \label{MIT}
  \and
  Caltech, Pasadena, CA 91125, USA
  \label{Caltech}
}

\abstract{%
Using the NIKA2 dual band millimeter camera installed on the IRAM30m telescope, 
we have mapped a relatively large field ($\sim70$~arcmin$^2$) in the direction of 
the star GJ526 to investigate the nature of the
sources found with the MAMBO camera at 1.2~mm ten years earlier. 
We have found that they must be  dust-obscured galaxies (SMGs) in the background beyond the star.
The new NIKA2 map at 1.15~mm reveals additional sources and, in fact, an overdensity 
of SMGs predominantly  distributed along a filament-like structure in projection on the sky 
across the whole observed field.
We speculate this might be a cosmic filament at high redshift as revealed  in cosmological hydrodynamical simulations.
Measurement of spectroscopic redshifts of the SMGs in the candidate filament 
is required now for a definitive confirmation of the nature of the structure.
}
\maketitle
\section{Introduction}
\label{intro}

Surveys at submillimeter and millimeter-wavelengths
have revealed a dust-obscured population of  galaxies at high redshift \citep{Smai97, Hugh98}.
These  massive SubMillimeter Galaxies (SMGs) are  forming stars at prodigious rates, sometimes
exceeding  1000~$M_{\odot}$/yr, with correspondingly bright far infrared luminosities larger than $10^{12}\,L_{\odot}$
\citep{Stra16}. SMGs are enshrouded in dust and so have only faint or undetectable
optical counterparts but contribute to  the far infrared cosmic background emitting approximately half
of all of the light in the universe \citep{Puge96,Dole06}.
Theories predict that galaxy formation preferentially occurs along large-scale
dark matter filamentary or sheetlike overdense regions that are related to initial fluctuations
of the density field in the primordial universe \citep{Bond96}
and formed the {\it cosmic web} as depicted by cosmological hydrodynamical simulations.
It is thought that star formation and galaxy growth at an early stage of cosmic evolution are critically 
tied to this network of dark matter filaments.
A large fraction of the neutral hydrogen from the inter
galactic medium falls into them and constantly replenishes
the gas reservoirs of galaxies through cold gas accretion ;
the cold gas streams along the filaments, survives the shocks at the virial
radii of the dark matter halos and infalls onto the galaxies to fuel their star formation and growth
\citep{Kere05, Deke09}. However, definitive observational confirmation of
cold gas accretion  along filaments of the {\it cosmic web} is actively sought.

A region of the sky several hundreds of arcseconds in size corresponds to an area of several cMpc 
at high reshift (scaling factor cMpc$/''$ does not depends strongly on $z$ at high $z$ in standard cosmology) that  can be mapped 
efficiently at millimeter wavelengths using a modern camera such as NIKA2 on the IRAM30m radiotelescope 
\cite{Monf11,Cata14,Calv16,Bour16,Adam18,Pero20}.
SMGs detected in such a field  must be embedded in filaments according to the current paradigm 
and should be signposts of the {\it cosmic web} in the distant universe.
We present NIKA2 data that are a contribution to this endeavour due to serendipity.  

In 2007, the field in the  direction of the nearby star GJ526 was mapped in  1.2\,mm continuum waveband
using the MAMBO camera installed on the IRAM 30-meter radiotelescope at the time and revealed a quasi-alignment of the star and
 five compact sources  in projection on the sky. It was not possible  to decisively conclude whether
these sources were background SMGs or a large, clumpy circumstellar disk seen edge-on around
this  star \citep{Lest09}. As cautiously stated in this paper, only future observations could definitely decide since these sources 
would stay fixed if in the background or move at the same rate and in the same direction as the star if clumps of a circumstellar disk.  
So a possible displacement as large as 22.6$''$ in the south-west direction according  
to the star proper motion after ten years in 2017 motivated  us to observe 
this field with the NIKA2 camera to determine the origin of these sources.

In this contribution, we report our finding that the five MAMBO sources did not move over this 10 year period 
and so they must be SMGs in the distant background beyond the star. It was argued in \citep{Lest09} that this was
statistically unlikely in assuming a mean number surface density of SMGs but we are now led to believe that the  
GJ526 field presents instead an overdensity of SMGs.  
In this paper, we discuss the remarkable spatial distribution of these sources in the 1.15\,mm NIKA2 map
along a filament-like structure  reminiscent of the {\it cosmic web} revealed in the cosmological
hydrodynamical simulations.

\begin{figure}[h!]
   \centering
   \includegraphics[width=7cm, scale=0.8]{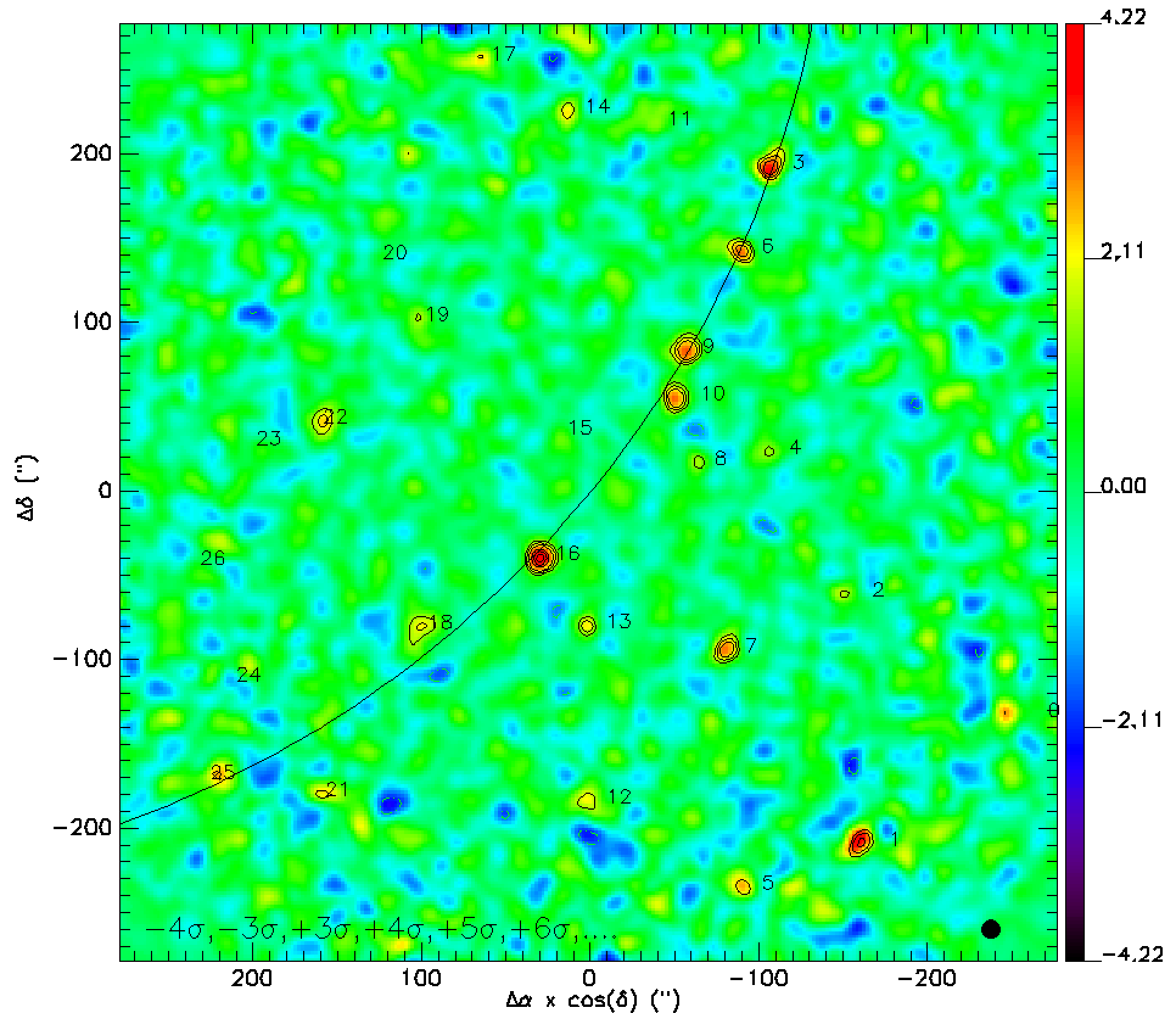} \includegraphics[width=5.5cm]{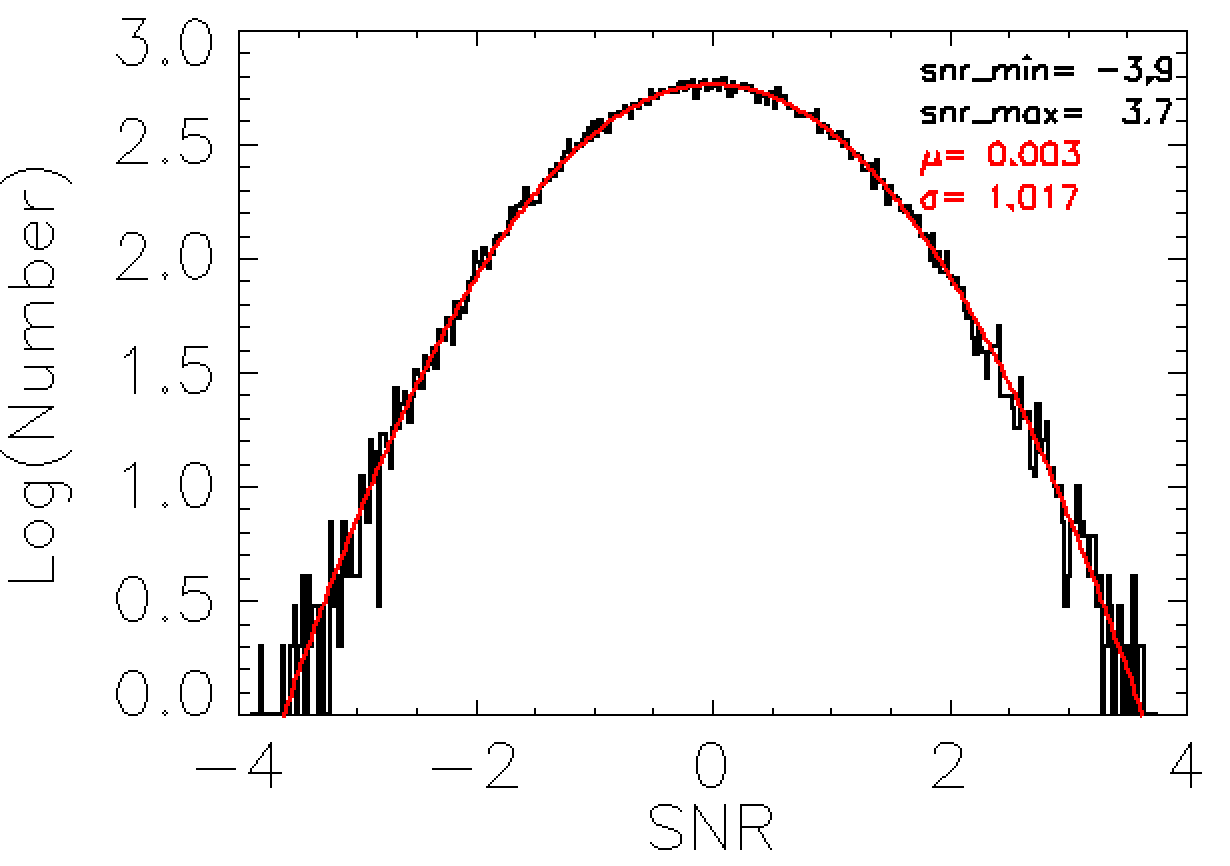}
   \caption{{\sl Left : } GJ526 NIKA2 map at 1.15mm (color scale : mJy/beam, contours : 
signal-to-noise ratio). Curve traces the filament-like structure identified across the map.
Beam size is $11.2''$ (black filled circle at brc). 
Map center (J2000) : 13h45m44.52s and $+14^{\circ}53'20.6''$. {\sl Right : } Signal-to-noise ratio histogramm 
of the jack-knifed map at 1.15~mm (in red, Gaussian distribution with zero mean and $\sigma$=1 for comparison).}
\label{fig:1mm}
\end{figure}

\section{NIKA2 observations and the 1.15~mm map}
\label{sec-1}

The NIKA2 observations  were obtained during the science verification
observations of the NIKA2 camera on 2017 april 24, 2018 february 14, 17, 18, 19, 20
and 2018 march 14, in average weather conditions (atmospheric opacity between 0.20 and 0.45 at 225~GHz) and at elevation
between 33$^{\circ}$ and 67$^{\circ}$.  The observing strategy
consisted of a series of $9' \times 4.5'$ on-the-fly scans with a scanning
speed of 40''/sec along two orthogonal orientations ($+60^{\circ}$, $+150^{\circ}$) in the AZ-EL
coordinate system at the telescope to minimise residual striping patterns in the maps.
These observations are split into 167 scans that correspond to a total of 9.3 hours of observations on source.
The time ordered information (TOI) of each detector was processed in order to
remove the contribution of atmospheric and electronic correlated
noises and to obtain the maps at 1.15~mm and 2.0~mm following the
procedure described in \citep{Adam14}  and Ponthieu et al. (in preparation). 
Point sources are detected by matching the maps with a Gaussian Point-Spread-Function.
Position accuracy is better than 3~arcseconds (for each coordinate) for sources above $4\sigma$ and  
the absolute photometric accuracy is $\sim$15~\% \citep{Pero20}.

In the limited space of this presentation, we shall discuss only the NIKA2 map at 1.15~mm and
plan a complete presentation including the NIKA2 map at 2.0~mm in a forthcoming paper.
We present the resulting 1.15~mm NIKA2 map in Fig.\,\ref{fig:1mm} and  astrometry and photometry  
of the ten unambiguously detected 
sources with a signal-to-noise ratio  (SNR) larger than 4 in Table\,\ref{tab:Tab_resul_1mm}.  
In order to determine this detection threshold, we have analysed the map noise in producing a jack-knifed map to
compute SNRs in all pixels and found that their distribution are Gaussian, centered on zero and have a standard deviation of unity 
as expected for normal noise (see Fig.\,\ref{fig:1mm} right). In these conditions, no spurious sources with $SNR > 4$
 are statistically expected in the $500'' \times 500''$ map owing to the number of independant 
beams ($11.2''$ at 1.15~mm) in it. There are also sixteen additional sources with  $3~<~{\rm SNR}~<~4$ but three or four spurious 
positive sources are statistically expected at this significance level.  
Source IDs in  Fig.\,\ref{fig:1mm} and  Table\,\ref{tab:Tab_resul_1mm} are assigned to all  ${\rm SNR>3}$ sources.

\begin{table}
\centering
\caption{NIKA2 astrometry and photometry of unambiguously detected sources at 1.15~mm.}
\begin{tabular}{rcccccc}  
\hline \hline
        ID     & $\alpha_{2000}$ & $\delta_{2000}$  & $S_{260GHz}$      & $SNR$    \\
        H3     &  $({\rm h~~m~~s})$      &  ~$(^\circ~~'~~'')$ &   (mJy)      &  260GHz     \\
\hline \hline
          1 &     13:45:33.48  &     14:49:50.5  &  3.96$\pm$ 0.95 &  5.3   \\
          3 &     13:45:37.18  &     14:56:31.9  &  3.67$\pm$ 0.81 &  6.1   \\
          6 &     13:45:38.36  &     14:55:42.2  &  2.95$\pm$ 0.68 &  5.7   \\
          7 &     13:45:38.97  &     14:51:45.4  &  2.87$\pm$ 0.63 &  6.1    \\
          9 &     13:45:40.61  &     14:54:42.8  &  3.13$\pm$ 0.65 &  6.9   \\
         10 &     13:45:41.09  &     14:54:14.8  &  2.97$\pm$ 0.62 &  6.8    \\
         13 &     13:45:44.69  &     14:51:59.4  &  2.08$\pm$ 0.53 &  4.7     \\
         16 &     13:45:46.63  &     14:52:39.7  &  4.11$\pm$ 0.75 &  9.6   \\
         18 &     13:45:51.51  &     14:51:59.0  &  2.15$\pm$ 0.57 &  4.5      \\
         22 &     13:45:55.51  &     14:54: 1.1  &  2.40$\pm$ 0.63 &  4.6    \\
\hline \hline
\end{tabular}
\label{tab:Tab_resul_1mm}   
\end{table}




We found that four  of the five MAMBO sources (see section GJ526 in Table~2 of \cite{Lest09})
have been detected anew in the NIKA2 map at 1.15~mm and
are identified as ID~7, 9, 10, 16 in Table~\ref{tab:Tab_resul_1mm} herein. The fifth MAMBO source $-$ MM134543+145317, a 4.3$\sigma$ detection
 in \cite{Lest09} $-$ has been searched in the  1.15~mm NIKA2 map and identified but only at a level of 2.8$\sigma$.  
Nonetheless, such a difference in SNR between the NIKA2 and MAMBO observations is statistically expected.
As already mentionned in the introduction, our main finding is that all five MAMBO sources did not move between 2007 and 2017 and so they must be SMGs in the distant background.

The MAMBO flux densities given at 250~GHz  in Table~2 of \cite{Lest09}
are comparable with the NIKA2 flux densities in Table~\ref{tab:Tab_resul_1mm} herein
to better than 2.4 times the quadratically combined uncertainties of the
two  measurements. In Table~\ref{tab:Tab_resul_1mm}, the flux density uncertainties include
both statistical and systematic photometric uncertainties (this latter one is $\sim$15\% \citep{Pero20}). 
 

\section{Filament-like structure in the GJ526 field}\label{overdensity}

As stated above,  there are ten sources unambiguously detected (SNR$>4$) above 2~mJy  at 1.15~mm in the whole NIKA2 map. 
This can be usefully compared  to the cumulative source number counts in other fields at a similar wavelength.
To make this comparison meaningful, we shall restrict the map to its central region within
a radial distance of $150''$ where  
the noise rms is uniform at the 10\% level. In this region,
there are six sources (ID~7, 9, 10, 13, 16, 18) that are unambiguously detected above 2~mJy at  1.15\,mm. 
Combination of several single dish studies of the COSMOS field, the Lockman Hole North field and the GOODS field, 
provides the cumulative source number count  $N(S_{1.2mm}>2.0{\rm mJy})=500$~sources/deg$^2$  \citep{Lind11}. 
The Poisson probability that six sources be found within $r < 150''$ 
given this mean surface number density  is as low as  3.7\%. 
However, the GJ526 field    
we picked for our NIKA2 observations is one out of  50 similar-sized fields targeted in the initial MAMBO survey 
to search for circumstellar disks \cite{Lest09}. So this overdensity in the GJ526 field  might very well 
be a statistical fluctuation in the whole MAMBO survey.
We turn now to the more intriguing spatial distribution of these SMGs along a filament-like structure. 



It is remarkable that seven SMGs (IDs 3, 6, 9, 10, 16, 18, 25) are spatially distributed along a slightly curved 
filament-like structure in projection on the sky as highlighted in Fig.\,\ref{fig:1mm} with an arc (solid line).  
The first six sources are unambiguously detected  and reported in Table~\ref{tab:Tab_resul_1mm} while the last one (ID~25) is only 
a possible $3.3\sigma$  detection with a 1.15~mm flux density of $2.61 \pm 0.79$~mJy.  
We can estimate the probability that such a quasi-alignment occurs in using
the average source number surface density $N(S_{1.2mm}>2.0$mJy)=500~sources/degree$^2$, noted $\mu$ below,
in small, consecutive sectors with a deviation angle $\delta$ from strict alignment of $ \pm 10^{\circ}$
(see sketch in Fig.\,\ref{fig:sketch}). 
Then, for a triplet of sources, the Poisson probability for the third source not to deviate 
more than $\delta$  from  alignment with the first two sources is $\mu \times \pi r^2 \times 2 \delta/360^{\circ}$, 
where $r$ is the distance between the second  and the third sources. 

\begin{minipage}[]{.46\linewidth}
\centering\raisebox{\dimexpr \topskip-\height}{ 
\includegraphics[scale=0.7]{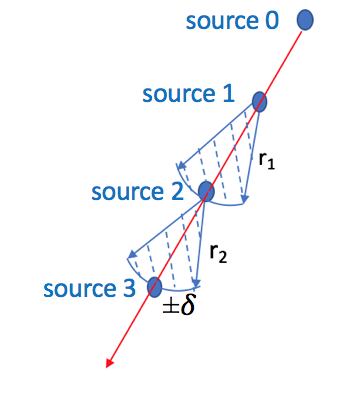}}
\captionof{figure}{Estimation of chance alignment of sources}
  \label{fig:sketch}
  \end{minipage}  \hfill
   \begin{minipage}[]{.46\linewidth}
As sketched in Fig.\,\ref{fig:sketch}, applying  
this formula for the first triplet in the NIKA2 map (source ID's~3, 6, 9), the probability of 
occurence of quasi-alignement within $\pm 10^{\circ}$ is   
$\rm{500~sources/deg}^2 \times \pi (70''/3600'')^2 \times 20^{\circ}/360^{\circ}=3.3\%$ where $r_1=70''$ is the
distance between source IDs 6 and 9. 
Similarly, the probability is 0.6\% for the second triplet (ID's~6, 9, 10) with $r_2=30''$,
 it is 11\% for the third triplet (IDs 9, 10, 16) with $r_3=130''$, it is 4.3\% for the fourth triplet (ID's~10, 16, 18) with
$r_4=80''$, and it is 15\% for the fifth triplet (ID's~16, 18, 25) with $r_5=150''$. 
  The combined probability  ($\prod\limits_{i=1}^{5}  \mu \pi r_i^2 2\delta/360^{\circ}$) 
is then as low as $1.4~10^{-7}$ and so it is very  unlikely that this
\end{minipage}
filament-like structure has occured by chance, 
even in accounting for the fact that 50 similar-sized fields have been observed in the initial MAMBO survey. 
So this apparent structure in the NIKA2 map is statistically significant. 

This filament-like structure stretches across the whole map which is $500''$ in size. So, if taken at the mean redshift 2.5 
of the B\'ethermin et al's SMG model \citep{Beth15}, it extends over $\sim$4~cMpc in 
using the scaling factor 8.3kpc/$''$ that does not depend strongly on redshift between 1 and 6 in standard cosmology \footnote{https://irsa.ipac.caltech.edu/Missions/herschel.html}.
In addition, the angular separations $r_i$  specified above between these seven SMGs correspond
 to linear separations comprised between $\sim$0.25~cMpc and $\sim$1.25~cMpc in using the same redshift and  scaling factor. This extent and
these separations are comparable with the spatial distribution of dark matter haloes hosting galaxies in the high resolution 
cosmological simulation {\it NewHorizon} 
\cite{Dubo21} for instance. This match between observations and simulations 
suggests that the structure in the NIKA2 map may be a cosmic filament.  A complete presentation of this hypothesis and
additional evidence, noticeably the luminosities $L_{FIR} > 10^{12}$~L$_{\odot}$ 
of these seven SMGs estimated with the 2.0~mm NIKA2 map 
and supplemental data from Herschel/SPIRE, will be published elsewhere.

\section{Conclusion}

In the 1.15~mm NIKA2 map of the GJ526 field presented here, there is an overdensity of SMGs 
characterized by an apparent filamentary structure in projection on the sky that is unlikely chance occurence and is 
 visually reminiscent of a cosmic filament as cosmological simulations have shown
({\it Illustris} \cite{Spri18}, {\it NewHorizon} \cite{Dubo21}, {\it Uchuu} \cite{Ishi21}). However, confirmation 
of such a filament requires now measurements of spectroscopic redshifts of the milliJansky unlensed SMGs in using
 molecular and 
atomic lines  with the most sensitive arrays such as NOEMA and ALMA. Also, comparison of our observed structure
to the most recent cosmological simulations listed above is presently hampered by the fact they do not 
track specifically the rapidly star forming galaxies (SMGs). In fact,
the less recent GADGET-2 simulation \cite{Dave10} was designed specifically to track
 massive galaxies with high star formation rates ($>280$~M$_{\odot}$/yr) but the study does not present 
their spatial distribution.
Specific tools to search cosmological simulations to better isolate SMGs are needed and complementary to the 
high-speed and deep mapping capability attained with  NIKA2.

While  galaxy surveys in the optical domain have already revealed the distinctive pattern of voids, sheets, filaments and nodes
characterising the structure of the universe at low redshift, spatial distribution of SMGs
have the potential  to reveal similar structure but at high redshift. 

\section*{Acknowledgements}
{This work is based on observations carried out during the science verification program of NIKA2
on the IRAM 30m telescope. IRAM is supported by INSU/CNRS (France), MPG (Germany) and IGN (Spain).
We are grateful to the IRAM staff for their dedicated support. This research has made use of the SIMBAD database,
operated at CDS, Strasbourg, France.
This research has made use of the NASA/IPAC Extragalactic Database (NED),
which is operated by the Jet Propulsion Laboratory, California Institute of Technology,
under contract with the National Aeronautics and Space Administration.
}

%
%
%


\end{document}